\PassOptionsToPackage{unicode}{hyperref}
\PassOptionsToPackage{hyphens}{url}
\documentclass[
]{article}
\usepackage{xcolor}
\usepackage[margin=1in]{geometry}
\usepackage{amsmath,amssymb}
\usepackage{iftex}
\ifPDFTeX
  \usepackage[T1]{fontenc}
  \usepackage[utf8]{inputenc}
  \usepackage{textcomp} 
\else 
  \usepackage{unicode-math} 
  \defaultfontfeatures{Scale=MatchLowercase}
  \defaultfontfeatures[\rmfamily]{Ligatures=TeX,Scale=1}
\fi
\usepackage{lmodern}
\ifPDFTeX\else
\fi
\IfFileExists{upquote.sty}{\usepackage{upquote}}{}
\IfFileExists{microtype.sty}{
  \usepackage[]{microtype}
  \UseMicrotypeSet[protrusion]{basicmath} 
}{}
\makeatletter
\@ifundefined{KOMAClassName}{
  \IfFileExists{parskip.sty}{%
    \usepackage{parskip}
  }{
    \setlength{\parindent}{0pt}
    \setlength{\parskip}{6pt plus 2pt minus 1pt}}
}{
  \KOMAoptions{parskip=half}}
\makeatother
\usepackage{graphicx}
\makeatletter
\newsavebox\pandoc@box
\newcommand*\pandocbounded[1]{
  \sbox\pandoc@box{#1}%
  \Gscale@div\@tempa{\textheight}{\dimexpr\ht\pandoc@box+\dp\pandoc@box\relax}%
  \Gscale@div\@tempb{\linewidth}{\wd\pandoc@box}%
  \ifdim\@tempb\p@<\@tempa\p@\let\@tempa\@tempb\fi
  \ifdim\@tempa\p@<\p@\scalebox{\@tempa}{\usebox\pandoc@box}%
  \else\usebox{\pandoc@box}%
  \fi%
}
\def\fps@figure{htbp}
\makeatother
\usepackage[]{natbib}
\usepackage{float}
\usepackage{booktabs}
\usepackage{amsmath}
\usepackage{amssymb}
\usepackage{longtable}
\usepackage{array}
\usepackage{bookmark}
\IfFileExists{xurl.sty}{\usepackage{xurl}}{} 
\hypersetup{
  pdftitle={From Noisy News Sentiment Scores to Interpretable Temporal Dynamics:~A Bayesian State-Space Model},
  pdfauthor={Ian Carbó Casals},
  hidelinks,
  pdfcreator={LaTeX via pandoc}}

\title{From Noisy News Sentiment Scores to Interpretable Temporal
Dynamics:~A Bayesian State-Space Model}
\author{Ian Carbó Casals}
\date{January 23, 2026}

\begin{document}
\maketitle

\begin{abstract}
Text-based sentiment indicators are widely used to monitor public and market mood, but weekly sentiment series are noisy by construction. A main reason is that the amount of relevant news changes over time and across categories. As a result, some weekly averages are based on many articles, while others rely on only a few. Existing approaches do not explicitly account for changes in data availability when measuring uncertainty. We present a Bayesian state-space framework that turns aggregated news sentiment into a smoothed time series with uncertainty. The model treats each weekly sentiment value as a noisy measurement of an underlying sentiment process, with observation uncertainty scaled by the effective information weight $n_{tj}$: when coverage is high, latent sentiment is anchored more strongly to the observed aggregate; when coverage is low, inference relies more on the latent dynamics and uncertainty increases. Using news data grouped into multiple categories, we find broadly similar latent dynamics across categories, while larger differences appear in observation noise. The framework is designed for descriptive monitoring and can be extended to other text sources where information availability varies over time.
\end{abstract}

 \newpage

\section{1. Introduction}\label{introduction}

Financial markets react quickly to new information. Because of this,
many studies use sentiment from text (news, social media, or messaging
apps) to understand investor mood, market risk, and possible price
movements. Today, modern Natural Language Processing (NLP) models can
score each message as positive or negative and can also classify
messages into categories
\citep{araci2019finbertfinancialsentimentanalysis, yin2019benchmarkingzeroshottextclassification, lewis2019bartdenoisingsequencetosequencepretraining}.
However, when we aggregate these AI scores over time, the resulting
sentiment series is often very noisy \citep{shapiro2022measuring}. It
can change a lot from week to week, not only because sentiment really
changes, but also because the amount of available information changes.
For example, a weekly sentiment value based on hundreds of messages is
much more reliable than a value based on only a few messages.

This problem is naturally framed as a latent-state estimation problem:
we observe noisy sentiment measurements, while the ``true'' sentiment is
a hidden process that evolves over time. A classic solution for this
type of problem is the Kalman Filter \citep{kalman1960new}, which
estimates a hidden state using a transition model (how the state
changes) and a measurement model (how observations are created). Later
work extends this idea to settings where observation noise is unknown or
changes over time, for example with variational Bayesian adaptive Kalman
filters, which estimate both the hidden state and time-varying noise
parameters \citep{huang2018adaptivekalmanfilter}. In finance, similar
dynamic models have been used for sentiment series, treating sentiment
indices as noisy measurements of an underlying latent sentiment process
and using filtering to separate long-term trends from short-term shocks
\citep{vassallo2021dynamic}. More recently, hybrid methods combine large
language models with Bayesian state-space models to improve forecasting
by using semantic information from text
\citep{cho2025llmintegratedbayesianstatespace}.

Even with these advances, an important gap remains for sentiment built
from irregular information streams, such as messaging platforms, social
media, or news feeds. In these sources, the number of relevant messages
can change strongly over time. Many existing models either assume a more
stable flow of information or try to learn time-varying noise in a
general way. But for sentiment data, the main driver of reliability is
often simple and observable: how much evidence do we have. If we ignore
this, we mix true sentiment changes with changes in data availability.
This is especially important because recent research shows that using
AI-generated variables as true data in econometric analysis can lead to
biased estimates and invalid inference due to measurement error
\citep{battaglia2024inference}. At the same time, recent surveys in
financial AI emphasize that uncertainty quantification (both aleatoric
and epistemic) is essential to make AI signals reliable for financial
decision-making \citep{eggen2025aleatoricepistemicuncertainty}.
Therefore, modeling measurement reliability is a first-order requirement
for sentiment monitoring from text.

In this paper, we introduce a hierarchical Bayesian state-space model to
build a stable and interpretable sentiment signal. We treat the observed
weekly sentiment for each category as a noisy measurement of a latent
sentiment state. The key idea is to scale observation noise using an
information density term \(n_{tj}\), defined as the total
category-relevance weight in week \(t\) (a weighted volume rather than a
raw message count). Unlike approaches that infer time-varying noise
through generic latent volatility \citep{huang2018adaptivekalmanfilter},
we use an observable, data-driven reliability measure based on category
relevance mass. When \(n_{tj}\) is high, the measurement is more precise
and observation noise is smaller; when \(n_{tj}\) is low, uncertainty
increases and the model relies more on latent temporal dynamics. A
Bayesian formulation provides full posterior distributions for latent
states and parameters, allowing us to report credible intervals and
quantify uncertainty in a principled way. This is essential for
monitoring noisy sentiment signals. Our approach follows the same
filtering principle as adaptive Kalman filtering
\citep{huang2018adaptivekalmanfilter} and sentiment filtering in finance
\citep{vassallo2021dynamic}, but it provides a more direct and
interpretable way to model changing measurement reliability in irregular
news streams.

Our contributions are threefold. First, we introduce a Bayesian
state-space model for category-level sentiment monitoring in which
observation noise is scaled by information density, linking data
availability to measurement uncertainty. Second, we use a hierarchical
structure to share information across categories, improving robustness
when some categories have few messages. Third, we provide an
uncertainty-aware framework for monitoring sentiment dynamics, which can
support more reliable downstream analysis and decision-making.

\section{2. Methods}\label{methods}

\subsection{2.1 Data and Temporal
Aggregation}\label{data-and-temporal-aggregation}

\subsubsection{2.1.1 News Collection and NLP
Scoring}\label{news-collection-and-nlp-scoring}

News articles were collected automatically from multiple public sources
using a custom Python pipeline. The data were stored in a relational
database and processed using pretrained natural language processing
(NLP) models to obtain sentiment and category relevance scores for each
article. The code used for data collection and NLP scoring, as well as
the analysis scripts, is publicly available (see Section 6 for data and
code availability).

Let \(X\) denote the text of a news article.

\paragraph{Sentiment.}\label{sentiment.}

We use the FinBERT model to estimate the sentiment of each article
\citep{araci2019finbertfinancialsentimentanalysis}. For a given text
\(X_i\), the model outputs probabilities associated with positive,
neutral, and negative sentiment:

\[
POS(X_i),NEU(X_i),NEG(X_i) \in (0,1)
\]

with \(POS(X_i) + NEU(X_i) + NEG(X_i) \approx 1\). We summarize these
outputs into a single sentiment score defined as:

\[
S(X_i) = POS(X_i) - NEG(X_i)
\]

which takes values in the interval \([-1, 1]\). Positive values indicate
predominantly positive sentiment, while negative values indicate
predominantly negative sentiment.

\paragraph{Category Tagging.}\label{category-tagging.}

To assign news articles to thematic categories, we use the zero-shot
classification model bart-large-mnli
\citep{lewis2019bartdenoisingsequencetosequencepretraining, yin2019benchmarkingzeroshottextclassification}.
Given a predefined set of \(m\) categories, the model returns a
relevance score for each category. For a given article \(i\) with text
\(X_i\), we denote by:

\[
C_{ij}(X_i) \in (0,1) \quad j = 1, \ldots, m
\]

the relevance score of category \(j\). For simplicity, we write
\(C_{ij}\) for \(C_{ij}(X_i)\) in the remainder of the paper. These
scores are interpreted as soft weights, allowing a single article to
contribute to multiple categories with different intensities.

\subsubsection{2.1.2 Construction of Sentiment
Scores}\label{construction-of-sentiment-scores}

Each article is assigned to a unique time window according to its
publication date. Time is divided into fixed windows of length
\(\Delta t\). In this study, we use weekly windows and set
\(\Delta t = 7\) days. Weekly windows are indexed as \(t=1,\dots,N\),
from the week starting \(\text{2023-12-31}\) to the week starting
\(\text{2025-11-30}\).

At the article level, the observed quantities are the sentiment scores
\(S_i\) and the category relevance scores \(C_{ij}\). In order to reduce
noise in the category assignments, we apply a threshold to the relevance
scores. Specifically, values below \(0.25\) are considered
non-informative and are set to zero:

\[
C_{ij} = 0 \quad \text{if } C_{ij} < 0.25
\]

After thresholding, articles with null vectors were excluded from the
analysis.

Using the processed category scores, we define the aggregated sentiment
score for time window \(t\) and category \(j\) as a weighted average
over all articles published in that interval. Let \(\text{news}_t\)
denote the set of articles published in time window \(t\). The
aggregated sentiment is defined as:

\[
y_{tj} = \frac{\sum_{i \in \text{news}_{t}} C_{ij} S_i}{\sum_{i \in \text{news}_t} C_{ij}}
\]

Finally, categories for which no articles remain after this procedure in
a large number of time windows are excluded from the analysis. Starting
from an initial set of eight categories, this filtering step results in
a final set of six categories used in the empirical analysis, which are
shown in Figure \(1\).

\subsubsection{2.1.3 Exploratory Analysis}\label{exploratory-analysis}

\begin{figure}[H]

{\centering \includegraphics[width=0.9\linewidth]{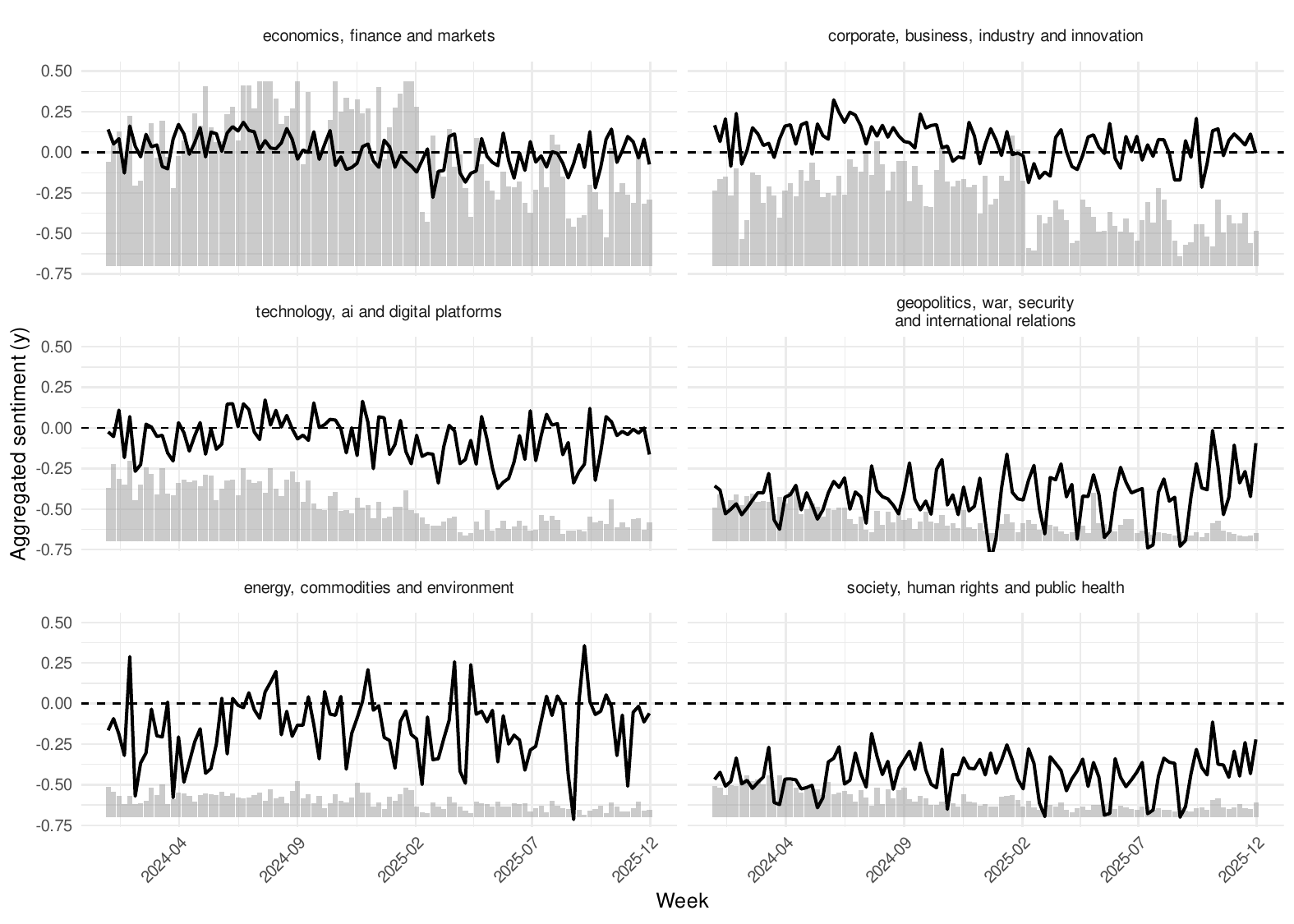} 

}

\caption{Aggregated sentiment $y_{tj}$ (line) and effective weight $n_{tj}$ (bars, globally rescaled) for each category.}\label{fig:fig_exploratory}
\end{figure}

Figure \(1\) shows the aggregated sentiment \(y_{tj}\) for each category
together with the effective weight \(n_{tj}\), defined as the total
category relevance mass within each time window (see Section 2.2.2 for
the formal definition). The bars represent \(n_{tj}\) and are globally
rescaled for visual comparison. After preprocessing, no time windows
with zero effective weight were observed in the \(6\) selected
categories.

The dataset contains a total of \(9,918\) news articles with positive
relevance for at least one category, corresponding to an average of
approximately \(99\) articles per week. Data availability varies over
time and, more clearly, across categories. Some categories are
consistently supported by a larger amount of category-related news,
while others exhibit much lower effective weights in many time windows.

This imbalance across categories motivates partial pooling and a
state-space model in which observation uncertainty is allowed to vary
with information density \(n_{tj}\).

\subsection{2.2 Bayesian State-Space Model for News
Sentiment}\label{bayesian-state-space-model-for-news-sentiment}

We model the observed aggregated sentiment \(y_{tj}\) as a noisy
measurement of an unobserved (latent) sentiment process \(x_{tj}\)
evolving over time \citep{gelman2013bayesian, durbin2012state}.

\subsubsection{2.2.1 Latent Sentiment
Dynamics}\label{latent-sentiment-dynamics}

For each category \(j\), the latent sentiment is modeled as an
autoregressive process of order one \(AR(1)\) around a long-run mean
\(\mu_j\):

\begin{flalign*}
&x_{1j} \sim \text{Normal}\big(\mu_j,\, \sigma_{\eta j}^2\big) \\
&x_{tj} \sim \text{Normal}\bigg((1-\theta_j)\mu_j + \theta_jx_{t-1,j},\, \sigma_{\eta j}^2\bigg) \quad t = 2,...,N 
\end{flalign*}

The persistence parameter \(\theta_j \in [0,1]\) controls how strongly
the current latent sentiment depends on the previous state. Values close
to \(1\) imply high persistence (slow changes), while values closer to
\(0\) imply faster mean reversion toward \(\mu_j\). The parameter
\(\sigma_{\eta j}\) controls the latent process variability.

\subsubsection{2.2.2 Observation Model and Effective
Weight}\label{observation-model-and-effective-weight}

The observed aggregated sentiment \(y_{tj}\) is modeled as a noisy
observation of \(x_{tj}\):

\[
y_{tj} \sim Normal\big(x_{tj},\, \sigma_j^2 / n_{tj}\big)
\]

We define \(x_{tj}\) as the latent state. Importantly, it represents the
underlying mean of the observed sentiment process (what we would measure
with complete coverage), not a different target outside this measurement
process (e.g., the sentiment value that would be obtained if we had
access to all relevant articles for category \(j\) in week \(t\) and
measured them without error).

We adopt a Gaussian observation model. Although sentiment is bounded in
\([-1,1]\), the posterior predictive probability of generating values
for \(y_{tj}\) outside \([-1,1]\) is negligible (see Appendix A.2, Table
2; \(\max_{tj} \ P(\text{out})<1.5\times10^{-3}\)).

Here \(\sigma_j\) is the observation noise scale for category \(j\). The
term \(n_{tj}\) represents the effective amount of category-related
information available in time window \(t\) for category \(j\), defined
as:

\[
n_{tj} = \sum_{i \in \text{news}_t} C_{ij}
\]

Intuitively, \(n_{tj}\) increases when more articles in window \(t\) are
strongly related to category \(j\) (high \(C_{ij}\)). The scaling
\(\sigma_j^2 / n_{tj}\) is motivated by the fact that \(y_{tj}\) is a
weighted average of many article-level sentiment scores. Under the
assumption that these articles provide roughly independent and diverse
evidence about category \(j\) in week \(t\), weighted averages become
more stable when they are computed from more effective evidence. For
this reason, we model measurement uncertainty as decreasing with
\(n_{tj}\): when \(n_{tj}\) is large, \(y_{tj}\) is a more precise
measurement of \(x_{tj}\); when \(n_{tj}\) is small, the model
downweights \(y_{tj}\) and relies more on the latent dynamics.

\subsubsection{2.2.3 Priors and Hierarchical
Structure}\label{priors-and-hierarchical-structure}

The long-run mean \(\mu_j\) is constrained to \([-1,1]\) using a
transformation:

\[
\mu_j^{aux} \sim \text{Beta}(1,1), \quad \mu_j = 2\mu_j^{aux} - 1
\]

To share information across categories while allowing category-specific
behavior, we place hierarchical priors on the persistence and noise
parameters.

Persistence. We model \(\theta_j\) on the logit scale:

\[
\theta_j^{aux} \sim \text{Normal}\big(\mu_\theta, \sigma_\theta^2\big), \quad \theta_j = \frac{1}{1+e^{-\theta_j^{aux}}}
\]

with hyperpriors \(\mu_\theta \sim \text{Normal}(0,1^2)\) and
\(\sigma_\theta \sim \text{Lognormal}(log(0.7),0.35^2)\).

Noise scales. We model standard deviations on the log scale:

\begin{flalign*}
&log(\sigma_j) \sim \text{Normal}\bigg( \mu_{log(\sigma)},\, \sigma_{log(\sigma)}^2 \bigg), \quad \sigma_j = e^{log(\sigma_j)} \\
&log(\sigma_{\eta j}) \sim \text{Normal}\bigg(\mu_{log(\sigma_\eta)},\, \sigma_{log(\sigma_\eta)}^2\bigg), \quad \sigma_{\eta j} = e^{log(\sigma_{\eta j})}
\end{flalign*}

The hyperpriors assigned are:

\begin{flalign*}
&\mu_{\log (\sigma)} \sim \text{Normal}(log(0.15),\,1^2), \quad \sigma_{\log (\sigma)} \sim \text{Lognormal}(log(0.5),0.35^2) \\
&\mu_{\log (\sigma_\eta)} \sim \text{Normal}(log(0.05),\,1^2), \quad \sigma_{\log (\sigma_\eta)} \sim \text{Lognormal}(log(0.5),0.35^2)
\end{flalign*}

This parameterization improves numerical stability in MCMC sampling and
guarantees positivity of \(\sigma_{\eta j}\) and \(\sigma_j\).

\section{3. Results}\label{results}

\subsection{3.1 Posterior Inference on Model
Parameters}\label{posterior-inference-on-model-parameters}

We analyze the posterior distributions of the main model parameters and
how they differ across categories. We focus on the persistence parameter
\(\theta_j\), the latent process scale \(\sigma_{\eta j}\), and the
observation noise scale \(\sigma_j\).

The results shown in this section are based on two estimations of the
same state-space model fitted to the same data, with identical
likelihood structure and priors. One estimation imposes full pooling
across categories, while the other allows category-specific parameters
through hierarchical priors.

In this section, black curves correspond to the pooled (simple) model,
where a single parameter is shared across all categories. Colored curves
correspond to the hierarchical model, where parameters are
category-specific but partially pooled through hierarchical priors.

\begin{figure}[H]

{\centering \includegraphics[width=0.75\linewidth]{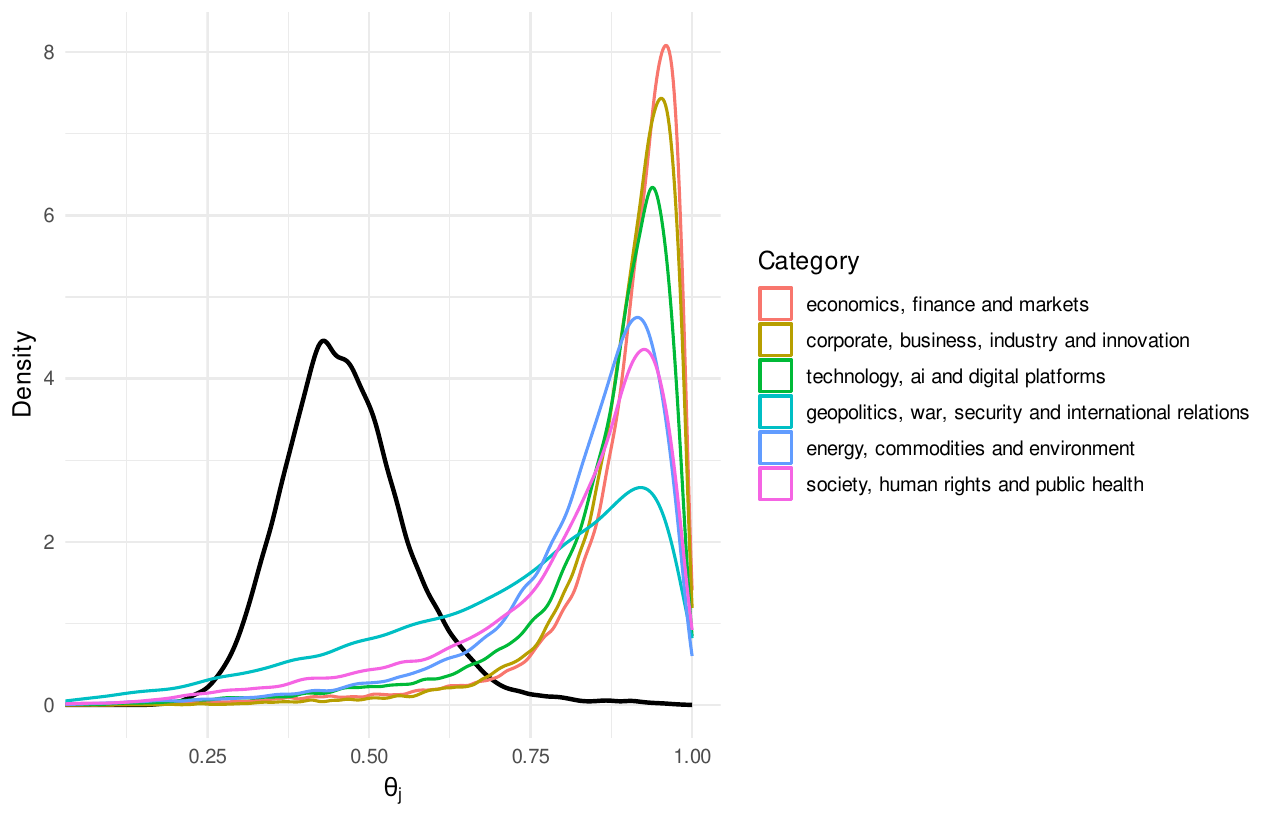} 

}

\caption{Posterior densities of the persistence parameter, pooled (black) vs category-specific (colors).}\label{fig:fig_theta}
\end{figure}

\subsubsection{3.1.1 Persistence of sentiment
dynamics}\label{persistence-of-sentiment-dynamics}

Figure \(2\) shows the posterior densities of the persistence parameter
\(\theta_j\). Compared to the pooled model, the hierarchical model shows
slightly different persistence levels across categories. However, these
differences are relatively small, and the category-specific posteriors
remain close to each other and centered around a common value of
\(\theta\). This indicates that most of the temporal persistence is
shared across categories, and that the main benefit of the hierarchical
model comes from modeling other sources of variation, rather than from
differences in persistence.

While the estimated persistence differs between the pooled and
hierarchical models, this does not indicate a contradiction. In the
pooled specification, a single persistence parameter must absorb most of
the temporal structure. In the hierarchical model, part of this
structure is instead captured by category-specific innovation and noise
terms. As a result, the persistence parameter mainly reflects a common
temporal component shared across categories.

\begin{figure}[H]

{\centering \includegraphics[width=0.75\linewidth]{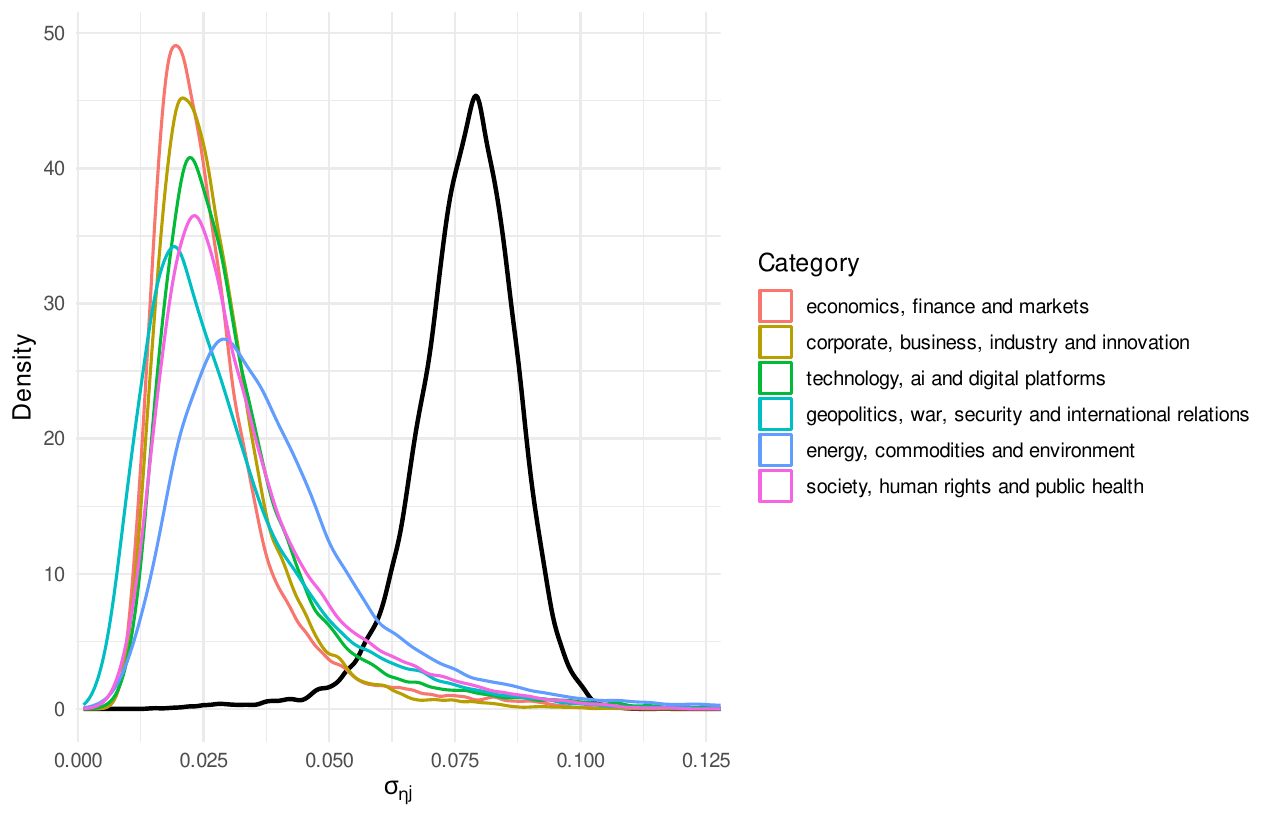} 

}

\caption{Posterior densities of the latent innovation parameter, pooled (black) vs category-specific (colors).}\label{fig:fig_sigma_eta}
\end{figure}

\subsubsection{3.1.2 Latent innovation
variability}\label{latent-innovation-variability}

Figure \(3\) presents the posterior densities of the latent innovation
standard deviation \(\sigma_{\eta j}\), which controls short-term
variability in the latent sentiment process \(x_{tj}\).

The hierarchical model shows small differences across categories.
Smaller values of \(\sigma_{\eta j}\) correspond to smoother latent
sentiment paths, while larger values allow for greater short-term
variation. Overall, the innovation scale remains similar across
categories, suggesting that short-term sentiment volatility is largely a
shared feature rather than strongly category-specific.

\begin{figure}[H]

{\centering \includegraphics[width=0.75\linewidth]{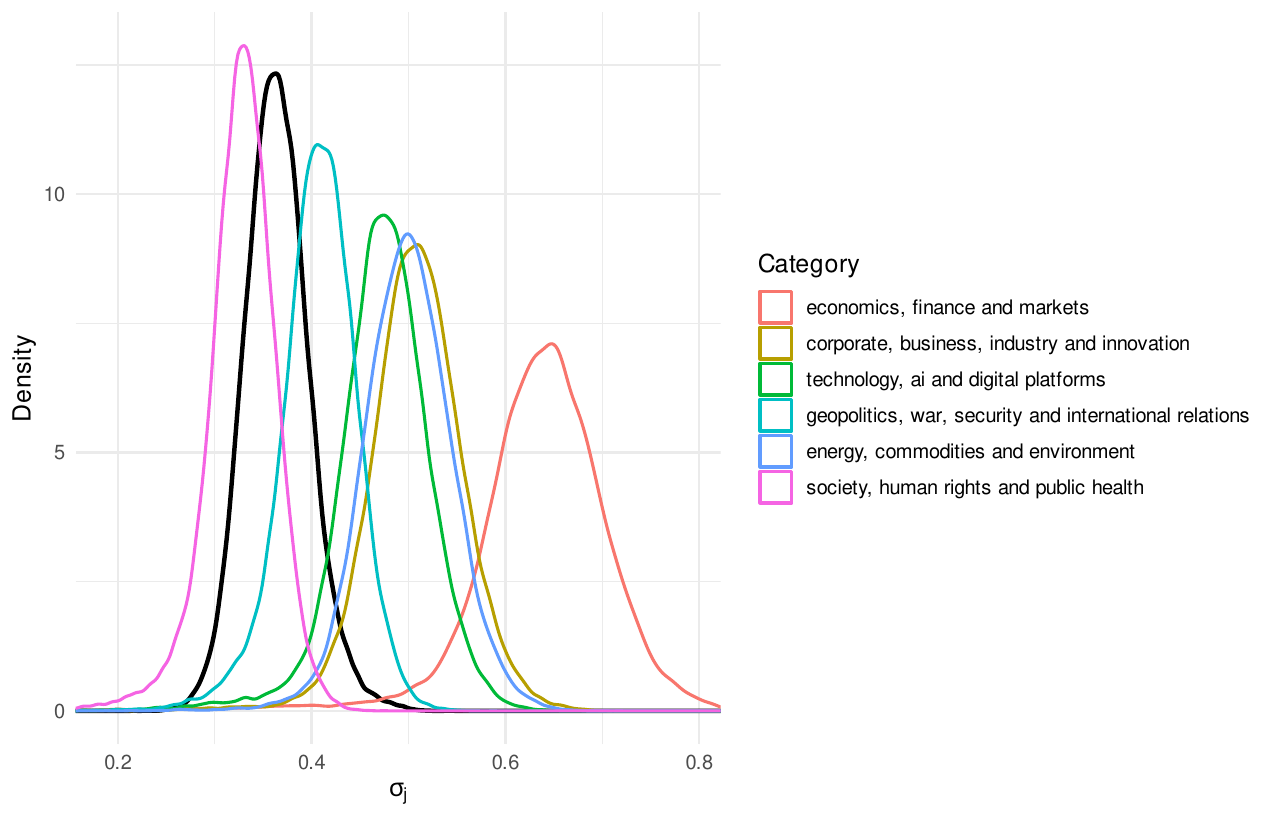} 

}

\caption{Posterior densities of the observation noise parameter, pooled (black) vs category-specific (colors).}\label{fig:fig_sigma}
\end{figure}

\subsubsection{3.1.3 Observation noise}\label{observation-noise}

Figure \(4\) shows the posterior densities of the observation noise
parameter \(\sigma_j\). This parameter measures how noisy the observed
aggregated sentiment \(y_{tj}\) is around the latent sentiment
\(x_{tj}\), after accounting for the effective category weight
\(n_{tj}\).

We find that observation noise differs across categories. This means
that the aggregated observed sentiment \(y_{tj}\) is a less precise
measurement of the latent state for some categories than for others,
likely reflecting differences in category ambiguity or variability in
news content.

An additional comparison re-estimates the same model on the same data,
allowing \(\sigma_j\) to vary by category while keeping \(\theta\) and
\(\sigma_\eta\) shared. In this setting, the posterior distributions of
\(\theta\) and \(\sigma_\eta\) closely match those obtained under the
fully hierarchical model. This shows that the differences observed
between the fully pooled and hierarchical models are driven primarily by
allowing for heterogeneity in observation noise across categories,
rather than by category-specific persistence or innovation dynamics.

\subsection{3.2 Observed and Latent
Sentiment}\label{observed-and-latent-sentiment}

\begin{figure}[H]

{\centering \includegraphics[width=0.95\linewidth]{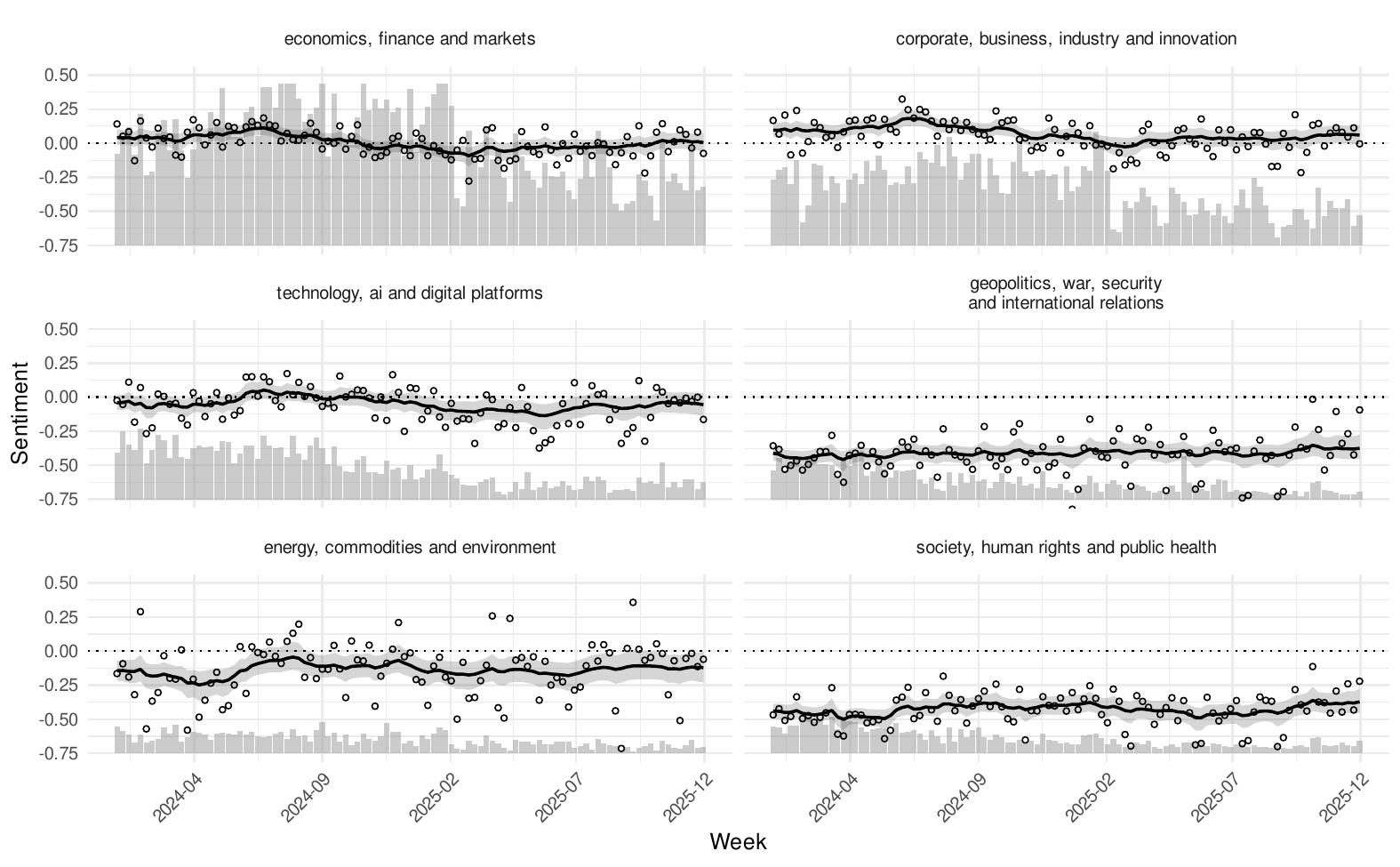} 

}

\caption{Observed $y_{tj}$ (points) and latent state sentiment $x_{tj}$ (solid line) with its 90\% credible interval.}\label{fig:fig_y_vs_x}
\end{figure}

Figure \(5\) compares the observed aggregated sentiment \(y_{tj}\)
(dots) with posterior estimates of the latent state \(x_{tj}\) across
categories. The observed series \(y_{tj}\) can change quickly from week
to week, while \(x_{tj}\) captures the more stable part of the sentiment
dynamics over time.

The observed series \(y_{tj}\) itself can be viewed as a simple
descriptive reference (direct aggregation) that does not model temporal
dependence or uncertainty. In particular, short-term fluctuations in
\(y_{tj}\) are more pronounced during periods of sparse coverage, making
it difficult to assess whether changes reflect signal or measurement
noise.

\subsection{3.3 Model Comparison: Heteroscedastic versus homoscedastic
observation
noise}\label{model-comparison-heteroscedastic-versus-homoscedastic-observation-noise}

To isolate the role of information density in the observation model, we
estimate two variants of the same state-space model on the same data,
with identical priors and hierarchical structure, that differ only in
the specification of the observation variance. This comparison provides
a direct validation of the core modeling assumption: posterior
uncertainty should increase mechanically as information density
\(n_{tj}\) decreases. In the proposed heteroscedastic model:

\[
y_{tj} \ |\ x_{tj} \sim N\big(x_{tj},\ \sigma_j^2/n_{tj}\big)
\]

which can be written as
\(y_{tj}=x_{tj}+(\sigma_j/\sqrt n_{tj})\epsilon_{tj}\) with
\(\epsilon_{tj} \sim N(0,1)\). This implies that lower \(n_{tj}\) should
lead to higher measurement noise and wider posterior uncertainty about
\(x_{tj}\). In a homoscedastic variant, we keep all other components
unchanged but use a constant observation variance:

\[
y_{tj} \ |\ x_{tj} \sim N\big(x_{tj},\ \sigma_j^2\big)
\]

Because \(x_{tj}\) is latent, we do not test this scaling through direct
residual standardization. Instead, we validate its observable
consequence on posterior uncertainty. Let \(Q_p(x_{tj})\) denote the
posterior p-quantile of \(x_{tj}\). We compute the 90\% posterior
interval width \(w_{tj}=Q_{0.95}(x_{tj})-Q_{0.05}(x_{tj})\) and regress
it on \(1/ \sqrt n_{tj}\), including category fixed effects and an
interaction with the homoscedastic variant. The regression output is
reported in Appendix A.1. In the heteroscedastic specification, a
one-unit increase in \(1/ \sqrt n_{tj}\) is associated with a \(0.098\)
increase in \(w_{tj}\), whereas in the homoscedastic variant the
corresponding marginal effect is \(0.041\). The difference in slopes is
statistically highly significant (interaction \(p<10^{-15}\)), showing
that the proposed scaling makes posterior uncertainty more responsive to
changes in information density.

\subsection{3.4 MCMC diagnostics}\label{mcmc-diagnostics}

Posterior inference was performed using Markov Chain Monte Carlo methods
implemented in JAGS \citep{plummer2003jags}, accessed through the
\texttt{r2jags} package in R.

We ran \(3\) MCMC chains of \(255,000\) iterations each, discarding the
first \(5,000\) as burn-in and retaining every \(25\)th draw, resulting
in \(10,000\) posterior samples per chain (\(30,000\) total).
Convergence was assessed using the \(\hat R\) diagnostic and visual
inspection of trace plots. For all main model parameters, \(\hat R\)
values were close to 1, and trace plots showed stable mixing across
chains. This suggests satisfactory convergence of the MCMC sampler.
Representative trace plots for selected parameters and categories are
reported in Appendix A.2, together with posterior predictive checks
(PPC).

\section{4. Discussion}\label{discussion}

Several limitations and modeling choices are worth noting. First, the
category relevance threshold used to define the weights \(C_{ij}\)
(0.25) is a pragmatic, dataset-specific choice. We selected it after
manually inspecting a small sample of articles and their category
scores, with the goal of cutting noise at a reasonable point for
aggregation. This threshold is therefore somewhat arbitrary and could be
tuned differently in other datasets; however, since it filters out
low-confidence assignments, moderate changes around this value should
have limited impact on the aggregated signals.

Second, the effective weight \(n_{tj}\) should be read as a simple and
interpretable measure of how much relevant news evidence is available in
each time window. In our setting, repeated coverage of the same story
across outlets is not necessarily undesirable: if the same news appears
multiple times, this may reflect that it is more important and more
widely reported. Under the assumption that sources are sufficiently
diverse, using \(n_{tj}\) as proposed provides a transparent way for the
model to express higher uncertainty when coverage is sparse. Future work
could make this step more explicit by measuring source diversity week by
week.

Third, we model the latent dynamics as an \(AR(1)\) process mainly for
interpretability and parsimony. This provides a directly interpretable
persistence parameter \(\theta_j\) and keeps the model identifiable with
limited time-series length per category. More flexible dynamics (e.g.,
higher-order autoregression) are possible, but would add complexity and
reduce transparency, which is not the main goal in descriptive
monitoring.

Finally, the sentiment and category scores depend on external NLP models
(here, FinBERT and related components), which may be imperfect. A
principled way to reduce reliance on a single model is to combine
multiple model outputs through probabilistic fusion (e.g.,
\citep{amirzadeh2025bayesiannetworkfusionlarge}); however, due to
limited resources we opted for a simpler pipeline to keep the example
simple. As an additional robustness extension, one could replace
Gaussian observation noise with a heavier-tailed alternative (e.g.,
Student-\(t\)), at the cost of extra parameters.

A natural extension of the framework is to account for systematic
differences in news exposure. In this paper, sentiment is measured from
the observed news flow, without modeling how different audiences may be
exposed to different subsets of news. Future work could incorporate
source-level or platform-level information to distinguish changes in
measured sentiment from changes in the composition of the news flow
(e.g., by weighting articles by source reach or accounting for
differences in how news spreads across platforms).

More broadly, the framework is not limited to news. It applies whenever
sentiment is derived from text and information availability changes over
time. By producing smoothed trajectories with credible intervals that
reflect data availability, the model supports descriptive monitoring and
quantitative comparisons across time and categories, including
applications such as product reviews, brand tracking, and campaign
monitoring.

\section{5. Conclusions}\label{conclusions}

This paper proposes a Bayesian state-space framework to analyze
news-based sentiment dynamics across multiple thematic categories. Using
aggregated sentiment scores extracted from news articles, the model
separates short-term noise from smoother underlying sentiment trends by
introducing a latent process that evolves over time. Unlike simple
aggregation approaches, the proposed method provides a principled way to
combine temporal smoothing with uncertainty quantification, while
remaining closely linked to the observed data.

The empirical results indicate that persistence parameters \(\theta_j\)
and short-term variability are similar across categories, with only
small differences between categories, while larger differences appear in
the observation noise. The effective weight \(n_{tj}\), which captures
how informative the observed sentiment is in each time window, plays a
key role in the model: when fewer relevant articles are available,
posterior uncertainty increases and the latent state relies more
strongly on temporal smoothing rather than on the observed data. This
mechanism allows the model to adapt naturally to imbalances in data
availability across time and categories.

Overall, the proposed framework is explicitly designed for descriptive
monitoring of news sentiment, prioritizing interpretability and
uncertainty quantification over forecasting accuracy, and it provides a
stable representation of sentiment over time.

\section{6. Data and Code
Availability}\label{data-and-code-availability}

To support reproducibility, the code used to collect and process the
news data (Telegram ingestion, FinBERT sentiment scoring, BART-MNLI
category relevance, and dataset export), as well as the analysis scripts
for the Bayesian state-space model, is available at
\texttt{https://github.com/Bailduke/bayesian-sentiment-state-space-model}
\citep{carbo2026bayesian_sentiment_code}. This paper corresponds to the
tagged GitHub release \texttt{v1.0.0}, which freezes the exact version
of the code used to generate the results reported in this manuscript.
\newpage

\section{Appendix}\label{appendix}

\subsection*{A.1 Information-density calibration}

This subsection reports the regression table for the validation
described in Section 4.2.2. For each category \(j\) and time \(t\), we
compute the 90\% posterior interval width of the latent state:

\[
w_{tj}=Q_{0.95}(x_{tj})-Q_{0.05}(x_{tj})
\] where \(Q_p(x_{tj})\) denotes the posterior p-quantile of \(x_{tj}\).
We then estimate the following linear model:

\[
w_{tj} = \beta_0 + \beta_1 \frac{1}{\sqrt n_{tj}} + \beta_2I_{homo}+\beta_3 I_{homo} \frac{1}{\sqrt n_{tj}} + \gamma_j + \epsilon_{tj}
\] where \(I_{homo}=1\) for the Homoscedastic variant and \(I_{homo}=0\)
for the Heteroscedastic model, and \(\gamma_j\) denotes category fixed
effects.

\begin{table}[!h]
\centering
\caption{\label{tab:width_reg_table}Linear model for the 90\% posterior interval width $w_{tj}=Q_{0.95}(x_{tj})-Q_{0.05}(x_{tj})$ as a function of $1/\sqrt{n_{tj}}$, with category fixed effects and an interaction with the homoscedastic variant.}
\centering
\resizebox{\ifdim\width>\linewidth\linewidth\else\width\fi}{!}{
\fontsize{9}{11}\selectfont
\begin{tabular}[t]{lrrrr}
\toprule
Term & Estimate & Std. Error & t value & p value\\
\midrule
(Intercept) & 0.0927 & 0.0012 & 74.97 & 0e+00\\
1/sqrt(n\_tj) & 0.0978 & 0.0046 & 21.33 & 0e+00\\
I\_homo & 0.0080 & 0.0016 & 5.09 & 4e-07\\
I\_homo × 1/sqrt(n\_tj) & -0.0566 & 0.0053 & -10.72 & 0e+00\\
\bottomrule
\end{tabular}}
\end{table}

\noindent\textit{Note:} Category fixed effects are included but not
shown for brevity. \(N=1212\), Adjusted \(R^2=0.8126\).

\subsection*{A.2 Model diagnostics}

\begin{figure}

{\centering \includegraphics[width=0.9\linewidth]{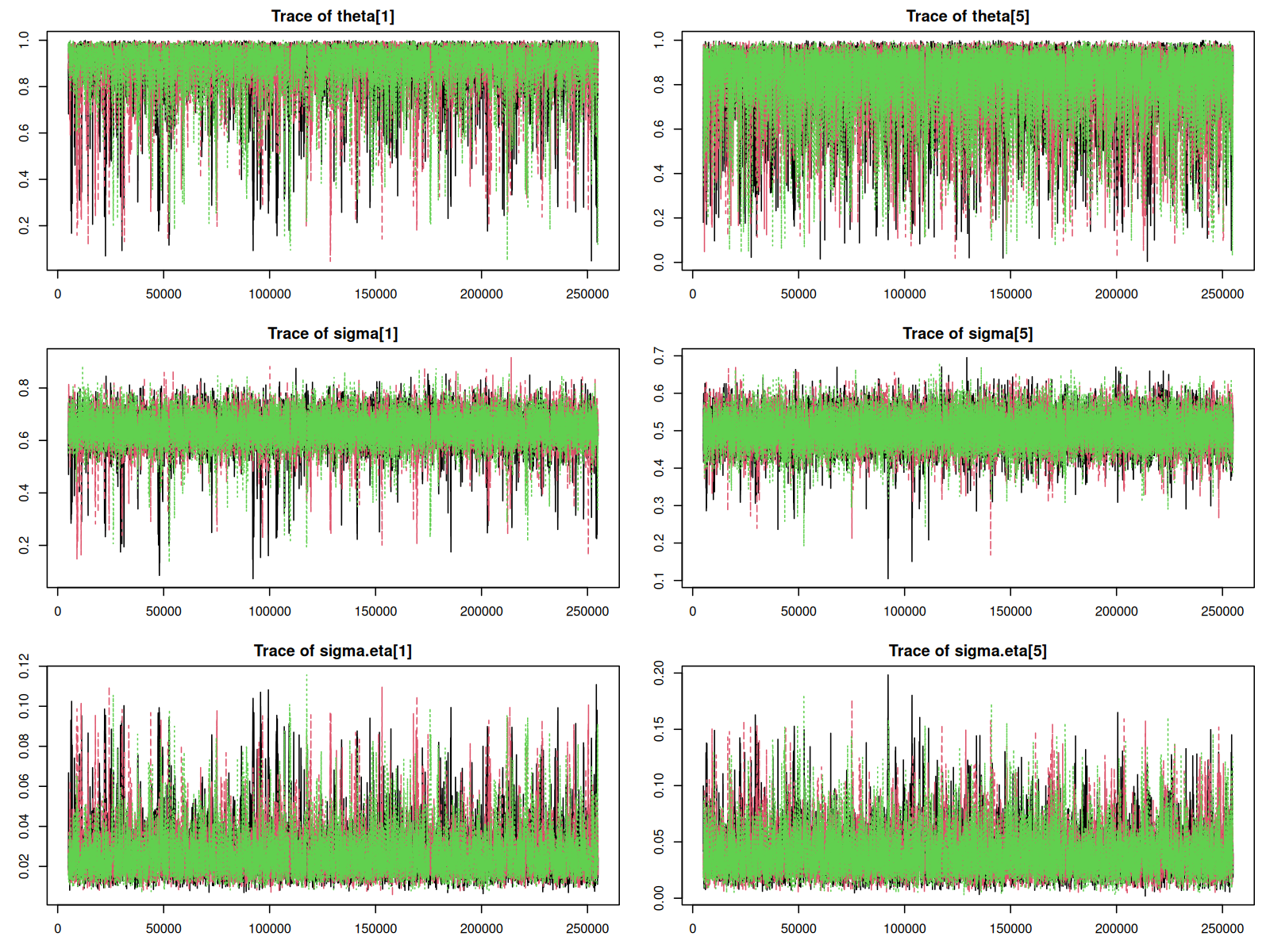} 

}

\caption{Trace plots for selected parameters in two representative categories (Economics, Finance and Markets; Energy, Commodities and Environment).}\label{fig:traceplots}
\end{figure}

Table \(2\) reports summary posterior predictive check (PPC) metrics for
the aggregated sentiment series. The \(80\)\% and \(95\)\% predictive
intervals cover the observed data at roughly the expected rates,
although the \(80\)\% intervals are a bit too wide. The posterior
predictive probability of generating values outside the valid sentiment
range \([-1,1]\) is negligible.

\begin{table}[!h]
\centering
\caption{\label{tab:ppc_table}Posterior predictive check (PPC) summary metrics for aggregated sentiment by category.}
\centering
\resizebox{\ifdim\width>\linewidth\linewidth\else\width\fi}{!}{
\fontsize{9}{11}\selectfont
\begin{tabular}[t]{>{\raggedright\arraybackslash}p{6.0cm}rrrr}
\toprule
Category & Cov 80\% & Cov 95\% & p (mean) & P(out)\\
\midrule
economics, finance and markets & 0.871 & 0.99 & 0.554 & 0.000000\\
corporate, business, industry and innovation & 0.881 & 0.98 & 0.790 & 0.000000\\
technology, ai and digital, platforms & 0.832 & 1.00 & 0.826 & 0.000000\\
geopolitics, war, security and international, relations & 0.851 & 0.98 & 0.706 & 0.000475\\
energy, commodities and environment & 0.901 & 0.96 & 0.574 & 0.001430\\
\addlinespace
society, human, rights and public, health & 0.861 & 0.99 & 0.612 & 0.000099\\
\bottomrule
\end{tabular}}
\end{table}

\newpage

\section{Acknowledgements}\label{acknowledgements}

I am grateful to Jordi Villà i Freixa, Full Professor at UVic--UCC and
group leader of the Computational Biochemistry and Biophysics Lab (CBBL)
at IRIS-CC, for his guidance and constructive feedback throughout the
development of this manuscript.

\renewcommand\refname{References}
\bibliography{references.bib}

@article{kalman1960new,
    author = {Kalman, R. E.},
    title = {A New Approach to Linear Filtering and Prediction Problems},
    journal = {Journal of Basic Engineering},
    volume = {82},
    number = {1},
    pages = {35-45},
    year = {1960},
    doi = {10.1115/1.3662552},
    url = {https://doi.org/10.1115/1.3662552},
    eprint = {https://asmedigitalcollection.asme.org/fluidsengineering/article-pdf/82/1/35/5518977/35_1.pdf},
}

@misc{cho2025llmintegratedbayesianstatespace,
      title={LLM-Integrated Bayesian State Space Models for Multimodal Time-Series Forecasting}, 
      author={Sungjun Cho and Changho Shin and Suenggwan Jo and Xinya Yan and Shourjo Aditya Chaudhuri and Frederic Sala},
      year={2025},
      eprint={2510.20952},
      archivePrefix={arXiv},
      primaryClass={cs.LG},
      url={https://arxiv.org/abs/2510.20952}, 
}

@techreport{battaglia2024inference,
  title={Inference for Regression with Variables Generated by AI or Machine Learning},
  author={Battaglia, Laura and Christensen, Timothy and Hansen, Stephen and Sacher, Szymon},
  year={2024},
  institution={Cowles Foundation for Research in Economics, Yale University},
  type={Cowles Foundation Discussion Paper},
  number={2421},
  url={https://elischolar.library.yale.edu/cowles-discussion-paper-series/2831}
}

@article{eggen2025aleatoricepistemicuncertainty,
author = {Eggen, Sivert and Espe, Tord Johan and Grude, Kristoffer and Risstad, Morten and Sandberg, Rickard},
title = {Financial Time Series Uncertainty: A Review of Probabilistic AI Applications},
journal = {Journal of Economic Surveys},
year = {2025},
url = {https://onlinelibrary.wiley.com/doi/abs/10.1111/joes.70018},
}

@article{huang2018adaptivekalmanfilter,
  author={Huang, Yulong and Zhang, Yonggang and Wu, Zhemin and Li, Ning and Chambers, Jonathon},
  journal={IEEE Transactions on Automatic Control}, 
  title={A Novel Adaptive Kalman Filter With Inaccurate Process and Measurement Noise Covariance Matrices}, 
  year={2018},
  volume={63},
  number={2},
  pages={594-601},
  doi={10.1109/TAC.2017.2730480},
  url={https://ieeexplore.ieee.org/document/8025799}
}

@phdthesis{vassallo2021dynamic,
  title={Dynamic models for financial and sentiment time series},
  author={Vassallo, Davide},
  year={2021},
  school={Scuola Normale Superiore},
  address={Pisa, Italy},
  url={https://hdl.handle.net/11384/109584}
}

@misc{amirzadeh2025bayesiannetworkfusionlarge,
      title={Bayesian Network Fusion of Large Language Models for Sentiment Analysis}, 
      author={Rasoul Amirzadeh and Dhananjay Thiruvady and Fatemeh Shiri},
      year={2025},
      eprint={2510.26484},
      archivePrefix={arXiv},
      primaryClass={cs.CL},
      url={https://arxiv.org/abs/2510.26484}, 
}

@misc{araci2019finbertfinancialsentimentanalysis,
      title={FinBERT: Financial Sentiment Analysis with Pre-trained Language Models}, 
      author={Dogu Araci},
      year={2019},
      eprint={1908.10063},
      archivePrefix={arXiv},
      primaryClass={cs.CL},
      url={https://arxiv.org/abs/1908.10063}, 
}

@misc{lewis2019bartdenoisingsequencetosequencepretraining,
      title={BART: Denoising Sequence-to-Sequence Pre-training for Natural Language Generation, Translation, and Comprehension}, 
      author={Mike Lewis and Yinhan Liu and Naman Goyal and Marjan Ghazvininejad and Abdelrahman Mohamed and Omer Levy and Ves Stoyanov and Luke Zettlemoyer},
      year={2019},
      eprint={1910.13461},
      archivePrefix={arXiv},
      primaryClass={cs.CL},
      url={https://arxiv.org/abs/1910.13461}, 
}

@misc{yin2019benchmarkingzeroshottextclassification,
      title={Benchmarking Zero-shot Text Classification: Datasets, Evaluation and Entailment Approach}, 
      author={Wenpeng Yin and Jamaal Hay and Dan Roth},
      year={2019},
      eprint={1909.00161},
      archivePrefix={arXiv},
      primaryClass={cs.CL},
      url={https://arxiv.org/abs/1909.00161}, 
}

@book{gelman2013bayesian,
  title={Bayesian Data Analysis},
  author={Gelman, Andrew and others},
  year={2013},
  publisher={CRC Press}
}

@book{durbin2012state,
  title={Time Series Analysis by State Space Methods},
  author={Durbin, James and Koopman, Siem Jan},
  year={2012},
  publisher={Oxford University Press}
}

@article{plummer2003jags,
  title={JAGS: A program for analysis of Bayesian graphical models using Gibbs sampling},
  author={Plummer, Martyn},
  journal={Proceedings of the 3rd International Workshop on Distributed Statistical Computing},
  year={2003}
}

@misc{carbo2026bayesian_sentiment_code,
  author       = {Ian {Carbó Casals}},
  title        = {bayesian-sentiment-state-space-model (v1.0.0)},
  year         = {2026},
  howpublished = {\url{https://github.com/Bailduke/bayesian-sentiment-state-space-model}},
  note         = {GitHub repository, release v1.0.0}
}

@article{shapiro2022measuring,
title = {Measuring news sentiment},
journal = {Journal of Econometrics},
author = {Adam Hale Shapiro and Moritz Sudhof and Daniel J. Wilson},
volume = {228},
number = {2},
pages = {221-243},
year = {2022},
issn = {0304-4076},
doi = {https://doi.org/10.1016/j.jeconom.2020.07.053},
url = {https://www.sciencedirect.com/science/article/pii/S0304407620303535}
}

\end{document}